\documentclass[letterpaper, fontsize=10.5pt]{scrartcl}	
\usepackage[margin=0.6in]{geometry}
\usepackage[T1]{fontenc}
\usepackage{fourier}
\usepackage[english]{babel}															
\usepackage[protrusion=true,expansion=true]{microtype}				
\usepackage{amsmath,amsfonts,amsthm,amssymb}
\usepackage{multicol}									
\usepackage[pdftex]{graphicx}														
\usepackage{url}
\usepackage{wrapfig}
\usepackage{subcaption}
\usepackage{booktabs}
\usepackage{siunitx}
\usepackage{enumitem}

\usepackage[normalem]{ulem}

\usepackage{pdfpages}
\usepackage{fancyhdr}
\newcommand{\jkfont}{
  \bfseries
  \color{orange}
}
\DeclareTextFontCommand{\jk}{\jkfont}

\newcommand{\sgfont}{
  \bfseries
  \color{magenta}
}
\DeclareTextFontCommand{\sg}{\sgfont}

\newcommand{\mafont}{
  \color{blue}
}
\DeclareTextFontCommand{\ma}{\mafont}

\newcommand{\gskfont}{
  \bfseries
  \color{red}
}
\DeclareTextFontCommand{\gsk}{\gskfont}

\newcommand{\editfont}{
  \bfseries
  \color{green}
}
\DeclareTextFontCommand{\edit}{\editfont}

\usepackage[numbers,sort&compress]{natbib}

\usepackage{titlesec, blindtext, color}
	\definecolor{light-blue}{rgb}{0.8,0.85,1}
	\definecolor{airforceblue}{rgb}{0.36, 0.54, 0.66}
	\definecolor{azure}{rgb}{0.0, 0.5, 1.0}
	\definecolor{bleudefrance}{rgb}{0.19, 0.55, 0.91}
	\definecolor{blue(munsell)}{rgb}{0.0, 0.5, 0.69}
	\definecolor{darkmidnightblue}{rgb}{0.0, 0.2, 0.4}
	\definecolor{steelblue}{rgb}{0.27, 0.51, 0.71}
	\definecolor{tealblue}{rgb}{0.21, 0.46, 0.53}
	\definecolor{yaleblue}{rgb}{0.06, 0.3, 0.57}
	\definecolor{applered}{rgb}{0.89, 0.02, 0.17}

\usepackage{sectsty}												
\allsectionsfont{\normalfont \large \scshape \color{yaleblue} \bfseries}	
\usepackage{setspace}

\usepackage[pdftex,unicode, implicit]{hyperref}
\hypersetup{
    unicode=true,          
    linktocpage=true,
    pdftoolbar=true,        
    pdfmenubar=true,        
    pdffitwindow=true,     
    pdfstartview={FitH},    
    pdfnewwindow=true,      
    colorlinks=true,       
    linkcolor=yaleblue,          
    citecolor=yaleblue,        
    filecolor=yaleblue,      
    urlcolor=yaleblue, 
   dvipdfm=true
}

\newcommand{\horrule}[1]{\rule{\linewidth}{#1}} 	

\title{ \normalfont 							
		\vspace{-1.025in} 	
		{\color{yaleblue}\horrule{2pt}} \\
		\Large	
		{\color{yaleblue}{\textsc{{\textbf{Solar Flare Energy Partitioning and Transport - The Impulsive Phase}}}}}\\ 
	    \small
		\textbf{Graham S. Kerr$^{1,2}$, Meriem Alaoui$^{1,2}$, Joel C. Allred$^{2}$, Nicolas H. Bian$^{3}$, Brian R. Dennis$^{2}$, A. Gordon Emslie$^{3}$, Lyndsay Fletcher$^{4,5}$, Silvina Guidoni$^{6}$, Laura A. Hayes$^{2}$, Gordon D. Holman$^{2~ (\mathrm{Emeritus})}$, Hugh S. Hudson$^{4}$, Judith T. Karpen$^{2}$, Adam F. Kowalski$^{7,8}$,\\Ryan O. Milligan$^{9}$, Vanessa Polito$^{10}$, Jiong Qiu$^{11}$, Daniel F. Ryan$^{6,2}$}\\
		\small
		\textsl{(1) Catholic U. of America, (2) NASA/GSFC, (3) Western Kentucky U., (4) U. of Glasgow, (5) U. of Oslo (6) American U., (7) U. of Colorado at Boulder, (8) National Solar Obs., (9) Queens U. Belfast, (10) Bay Area Environmental Research Inst., (11) Montana State U.}\\
		\small
		{\color{yaleblue}White Paper in response to NASA's \textsl{Heliophysics 2050 Workshop} Solicitation}\\
		\vspace{-0.05in}
		{\color{yaleblue}\horrule{2pt}}
		\vspace{-.25in} 
}
\newcommand{\apj}{ApJ}

\newcommand{\aap}{A\&A} 
\newcommand{\solphys}{Sol. Phys.}

\date{}

\begin{document}
\begin{spacing}{1}
\maketitle
\vspace{-1in}


\end{spacing}
\thispagestyle{empty}

\vspace{-0.1in}
\section{Overview}
\vspace{-0.15in}

Solar flares are a fundamental component of solar eruptive events (SEEs; along with solar energetic particles, SEPs, and coronal mass ejections, CMEs). Flares are the first component of the SEE to impact our atmosphere, which can set the stage for the arrival of the associated SEPs and CME. Magnetic reconnection drives SEEs by restructuring the solar coronal magnetic field, liberating a tremendous amount of energy which is partitioned into various physical manifestations: particle acceleration, mass and magnetic-field eruption, atmospheric heating, and the subsequent emission of radiation as solar flares. To explain and ultimately predict these geoeffective events, the heliophysics community requires a comprehensive understanding of the processes that transform and distribute stored magnetic energy into other forms, including the broadband radiative enhancement that characterises flares. 

This white paper discusses energy transport during the impulsive phase of flares. We discuss the gradual phase in a related white paper. Following \citep{2013arXiv1311.5243L} our approach to solar flare research from now until 2050 reflects the following basic philosophy: \textbf{{\color{yaleblue}(1)} to identify the sites of energy release and particle acceleration in the solar corona}; \textbf{{\color{yaleblue}(2)}} \textbf{to characterize the most energetically important components as they evolve in time and space}; and \textbf{{\color{yaleblue}(3)} to understand how 
energy is transported and dissipated, heating the Sun's atmosphere from photosphere to corona}. Observations should be made of these energy conversion 
sites before, during, and after the event, to characterize the magnetic fields, 
plasma density, temperature, flow velocities, wave fields, and the accelerated electrons and ions. These properties should be measured as close to the release site as possible, both in space and in time to minimize uncertainties due to propagation effects and temporal evolution. 

Our aim should be to ensure that models can reproduce fundamental and universal aspects of flares, and to improve the included physics where they cannot. Only then can we hope to model specific flare events. We must determine the acceleration mechanism and propagation properties of accelerated electrons. We must also push beyond the standard paradigm of energy transport via electron beams, particularly to address the roles of flare-accelerated ions, and Alfv\'en waves. To determine whether these models accurately represent the wide range of flare phenomena, we must confront them with high-quality observations. We outline key areas where progress would advance insight into flare physics substantially.  

\vspace{-0.175in}
\section{Nonthermal Electrons}
\vspace{-0.15in}
Nonthermal electrons are thought to be the primary means by which flare energy is transported through the solar atmosphere. Though great strides have been made, both observationally and theoretically, in our understanding of flare-accelerated electrons,  the following fundamental questions must be addressed: (1) where and on what time scales are electrons accelerated, and what is their energy distribution?; and (2) how do the accelerated electrons propagate from their coronal acceleration site to the chromosphere and into interplanetary space? 
\textbf{Electron acceleration.} Several flare-acceleration mechanisms have been proposed, which differ in the locations and physical processes involved.
Candidates include acceleration by electric fields at reconnection sites, contracting or merging magnetic islands along the current sheet, reconnection outflow jets, termination shocks or stochastic acceleration at the top of flare arcades, and large-scale and turbulent parallel electric fields, e.g., through kinetic Alfv\'en waves 
(see review by \citep{2011SSRv..159..357Z}). Establishing the locations (both coronal and footpoint sources), timescales, progression, and spectral characteristics of X-rays can discriminate among these acceleration mechanisms. 

\textbf{Electron propagation and energy deposition}. Efforts to explain the shape of X-ray spectra in terms of electron propagation \citep[][]{2008A&A...487..337B,2011ApJ...731..106S} or acceleration \citep{2008ApJ...683.1180G} have proven fruitful. Features such as strong breaks, though, prove difficult to explain by current propagation models \citep{2017ApJ...851...78A,2019SoPh..294..105A}. Many factors affect the transport of electrons from the acceleration region and, consequently, where and how much energy is deposited in the flaring atmosphere. For example, some propagation processes heat the corona, or result in trapping of electrons. 

These questions persist in part because present observations of hard X-ray (HXR) and microwave emission (MW), the signatures of electron acceleration, lack the dynamic range to access acceleration sites in the corona in the presence of the much brighter footpoint emission. Emissions at both locations are needed for an accurate characterisation of acceleration processes. To make progress discriminating between spectral features of the acceleration and propagation mechanisms, we require high spectral and temporal resolution ($\sim 1$~s) X-ray observations over an energy range at least up to 200 keV, with dynamic range ($>100:1$) sufficient to detect both regions. Imaging spectroscopy of the thermal plasma is needed to differentiate among mechanisms that yield heated coronal plasma: directly by the reconnection process; by the beam losses themselves (including Ohmic dissipation of the beam-neutralising return current); and by chromospheric ablation \citep{2016JGRA..12111667H}. At the same time we need models capable of self-consistently solving the transport equation including effects such as Coulomb collisions, return-current deceleration, magnetic mirroring, and synchrotron radiation \citep{2020arXiv200810671A}. These transport models should be used to interpret X-ray and microwave spectra, which provide fundamental aspects of the electron distribution, and to drive flare radiation-hydrodynamic simulations. Radio observations yield fundamental information about the the magnetic-field strength and the anisotropy of the electron distribution above $\gtrsim$200 keV, which are necessary to understand both acceleration and propagation mechanisms. The anisotropy can also be evaluated from X-ray polarimetry with improved sensitivity \citep{2017A&A...606A...2C}. In addition to the electron anisotropy, the directivity of the bremsstrahlung emission affects the observed photon spectrum and can potentially help determine the electron angular distribution \citep{2020arXiv200807849J}.

{The same reconnection sites that energize flares in SEEs also generate CMEs. Although flare-accelerated electrons and their radiative output have been studied intensely for decades, their upward-driven counterparts have received little attention. Those electrons penetrate the magnetic flux rope that comprises the CME, and may be released into the heliosphere as impulsive solar energetic particles \citep{2013ApJ...771...82M}. To separate the flux of downward and upward propagating electron populations in SEEs and understand their relationship, we need sensitive HXR and microwave imaging spectroscopy with high dynamic range.}

\vspace{-0.175in}
\section{Nonthermal Ions}
\vspace{-0.15in}
Most research into flare-accelerated particles has centered around nonthermal electrons. However, it is likely that ions (protons, and to a lesser extent $\alpha$ particles and other ions) are also accelerated in flares \citep{2000AIPC..522..401R,2009ApJ...698L.152S}, and 
might carry as much energy as the energetic electrons \citep{2012ApJ...759...71E,2017ApJ...836...17A}. Neglecting nonthermal ions in flare models means that {\color{yaleblue}\textsl{we are potentially ignoring up to half of the flare energy transported through the Sun's atmosphere.}} A significant theoretical and observational effort in the next few decades is crucial to plug this gap in our understanding of flares. 

Flare-accelerated protons have received less attention than electrons, in part due to the difficulty in detecting their nuclear $\gamma$-ray signatures compared to microwaves and X-rays produced by nonthermal electrons. Consequently observational constraints are lacking on the proton energy spectra needed to drive flare models, leading to a dearth of theoretical studies.  While we wait for more direct proton beam observations via $\gamma$-rays, we 
should use computational experiments to establish the possible range of nonthermal proton beam properties and study the effects of accelerated-ion transport through the solar atmosphere (e.g., energy loss mechanisms and scattering by wave-particle interactions \citep[e.g.,][]{1989ApJ...342..576T}). By comparing the resulting synthetic radiation to observations, we could determine whether proton beams can explain those observations that electron beams alone are unable to account for. For example, observations from the lower solar atmosphere, where proton beams can reach but electrons cannot \citep{1989SoPh..121..261N,1986ApJ...309..409T,2020arXiv200810671A}, can be compared to synthetic observables from simulations driven by electron, proton, or mixed-species beams. The most prominent example is white-light flares (WLFs), which are enhancements to the optical continuum whose origin has long been debated \citep[e.g.,][]{1989SoPh..121..261N,2014ApJ...783...98K}. There is evidence that WLFs are produced in the photosphere  \citep[][]{1989SoPh..121..261N,2014ApJ...783...98K,1982SoPh...80..113H,2013ApJ...776..123W}, though there is insufficient power carried by the highest energy electrons able to penetrate to the deepest atmospheric layers.  Other signatures of energy deposition into the deepest layers of the atmosphere have been identified in some large flares from spectroscopy \citep[e.g.,][]{1990ApJ...350..463M} and from sunquakes \citep[e.g.,][]{1998Natur.393..317K,2015SoPh..290.3163Z}. Clearly an alternative agent capable of penetrating to photospheric layers is required, and proton beams are a prime candidate. 

Directly establishing the fraction of the flare energy in accelerated ions, the distribution function of those ions, and their spatiotemporal relation to flare accelerated electrons is the best way to test acceleration models and drive flare heating models. For ions $>1$~MeV, these properties should be determined from $\gamma$-ray observations, for example of the $2.223$~MeV neutron-capture line, nuclear de-excitation lines between $1-10$~MeV, and the positron-annihilation line at $511$~keV. These observations can provide the flux and spectral distribution of the highest energy ions. Simultaneous observations of X-rays and $\gamma$-rays $<10$~keV to $>15$~MeV are required to address the relation between electrons and ions.

Flare-accelerated ions $<1$~MeV can be revealed by nonthermal Ly$\alpha$ (or Ly$\beta$) wing asymmetries that carry information about the flux, energy spectrum, and direction of the incident protons \citep{1976ApJ...208..618O}.  These emissions are produced by precipitating nonthermal protons that capture an electron from a neutral hydrogen atom (charge exchange), yielding nonthermal excited neutral hydrogen that subsequently emits a Doppler-shifted Ly$\alpha$ photon. Red-wing enhancements result from photons directly escaping upwards, and blue- wing enhancements result from downward-directed photons that are then scattered by neutral hydrogen. This spectral signature has not yet been observed on the Sun (owing largely to the paucity of Ly$\alpha$ flare observations), but has been observed in terrestrial aurora \citep{1989GeoRL..16..143I}.  A search for asymmetries in the wings of He~\textsc{ii} 304~\AA, caused by charge exchange with nonthermal $\alpha$ particles, failed to detect this effect \citep{2012ApJ...752...84H}, although previous theoretical investigations predicted that the effect should have been detectable. The estimated intensity of these signatures should be revised by flare codes capable of modelling the time-dependent, propagating proton/ion distribution with realistic ionization stratification.

\vspace{-0.175in}
\section{Alfv\'en Waves}
\vspace{-0.15in}
Magnetohydrodynamic (MHD) waves are undoubtedly produced in the corona during the large-scale restructuring of the magnetic field during flares \citep[e.g.,][]{2010PASJ...62..993K,2009ApJ...695.1151B,2012ApJ...760...81K,kumar2013,2017ApJ...844..149K}.  Downward propagating Alfv\'en waves may 
 heat the lower atmosphere via Joule dissipation of currents \citep[e.g.,][]{1982SoPh...80...99E} or ion-neutral friction \citep[e.g.,][]{2013ApJ...765...81R}. Energy transport by Alfv\'en waves is well established in the magnetosphere \citep[see reviews by][]{2009SSRv..142...73K,alfvenwave_review}, but their role in solar flare energy transport remains a mystery. A non-trivial amount of the energy could be partitioned into Alfv\'en waves, a possibility that demands exploration.

High-frequency ($f\gtrsim1$~Hz, to penetrate through the transition region \citep{2013ApJ...765...81R}) downward propagating Alfv\'en waves have received renewed attention as a potential vehicle to transport energy from the coronal energy release site to the chromosphere or deeper \citep{2008ApJ...675.1645F,2013ApJ...765...81R}, and even as a means to accelerate electrons locally in the chromosphere \citep{2008ApJ...675.1645F}. Several numerical experiments have explored their ability to heat the temperature-minimum region and the chromosphere, and have searched for observational signatures \citep{2016ApJ...818L..20R,2016ApJ...827..101K,2018ApJ...853..101R}. Those pioneering models are rather simplified, however, using (radiation) hydrodynamic codes to model the heating resulting from an approximated form of monochromatic Alfv\'en waves. 
 
Despite these approximations, such studies have demonstrated that Alfv\'en waves could efficiently heat the lower atmosphere, and produce mass flows consistent with observations; in addition, varying the wave parameters could heat different locations in the atmosphere, depending on the ionization stratification. However, the lack of definitive observational constraints on wave parameters has severely hampered efforts to make further progress. To determine whether downward propagating Alfv\'en waves play a significant role in flares, we need to focus on the following areas.

To address this question properly, radiative MHD simulations of flares must include an accurate NLTE chromosphere. However, a radiative MHD flare simulation that resolves the required spatial scales (down to meters in some situations) and incorporates energy transport via nonthermal particles is a challenge that may not be fully realized before 2050. To address this question, we may start with improvements to existing radiation hydrodynamic models, for example by implementing realistic wave spectra and propagation. On the longer term, we should aim to develop a full radiative MHD flare simulation capable of resolving Alfv\'en wave transport and dissipation and also of modeling the effects of nonthermal particles. The required spatial and temporal scales, and the range of physics processes to be included, make this  a very challenging task.

Observational efforts should focus on confirming the presence of these high-frequency waves in the chromosphere, 
measuring their properties, and determining where they originate and deposit their energy. This will require high cadence (sub-second) and high spatial resolution (less than 0.1$^{\prime\prime}$) spectroscopy of both the chromosphere and corona. Because the waves propagate deeper through the chromosphere on timescales of a few seconds, emission forming at varying depths could switch on sequentially, requiring broad temperature coverage of spectral lines. Measurements of nonthermal line broadening from spectral lines observed at a series of heights can be used to estimate the energy flux carried by the propagating waves \cite[e.g.,][]{1998A&A...339..208B,2017ApJ...845...98O}. The amount of energy carried by the waves depends on the magnetic-field perturbations, so the coronal magnetic-field fluctuations must be measured to the level of a few percent. The height dependence of the magnetic field and density stratification define the Alfv\'en speed, an important quantity for models to include accurately. Even a coarse observational sampling of the field strength at different heights would help, which could potentially be provided by the Daniel K. Inouye Solar Telescope (DKIST). 

The flare community would benefit here from collaboration with the magnetospheric community. The generation, propagation, and dissipation of Alfv\'en waves in the magnetosphere during reconnection is a well-established research area \citep{2009SSRv..142...73K,alfvenwave_review} that benefits from \textsl{in-situ} measurements. Although the characteristic plasma scales of the solar atmosphere and the magnetosphere differ substantially, for the most part, closer collaboration between flare and magnetospheric researchers would greatly advance our understanding of energy partitioning between particle acceleration and wave generation, the potential energy spectra of Alfv\'en waves, and pertinent wave-particle interactions.


\begin{spacing}{1}
   \vspace{-0.125in}
\begin{multicols}{2}
	\bibliography{Helio2050_GSKerr}
\end{multicols}
\end{spacing}
\end{document}